\documentclass[
    ,final   ]
  {aipproc}

\layoutstyle{8x11double}

\begin{document}

\title{The Subpulse Modulation Properties of Pulsars and its Frequency Dependence}

\classification{97.60.Gb}
\keywords      {Stars:pulsars:general --- Radiation Mechanisms: non-thermal}

\author{Patrick Weltevrede}{
  address={Australia Telescope National Facility, CSIRO, PO Box 76, Epping NSW 1710, Australia.\\ Email: Patrick.Weltevrede@atnf.csiro.au}
}

\author{<Ben Stappers>}{
  address={Stichting ASTRON, Postbus 2, 7990 AA Dwingeloo, The Netherlands.\\ Email: stappers@astron.nl}
}

\author{<R.T. Edwards>}{
 address={No longer working in astronomy}
}

\begin{abstract}
A large sample of about two hundred pulsars have been observed to
study their subpulse modulation at an observing wavelength of (when achievable) both 21
and 92 cm using the Westerbork Synthesis Radio Telescope. For 57
pulsars drifting subpulses are discovered for the first time and are
confirmed for many others. This leads to the conclusion that it could
well be that the drifting subpulse mechanism is an intrinsic property
of the emission mechanism itself, although for some pulsars it is
difficult or impossible to detect. It appears that the youngest
pulsars have the most disordered subpulses and the subpulses become
more and more organized into drifting subpulses as the pulsar
ages. Drifting subpulses are in general found at both frequencies and
the measured values of $P_3$ at the two frequencies are highly
correlated, showing the broadband nature of this phenomenon. 
Also the modulation indices measured at the two frequencies are
clearly correlated, although at 92 cm they are on average possibly higher. The correlations with the 
modulation indices are argued to be consistent with the picture in
which the radio emission is composed out of a drifting subpulse signal
plus a quasi-steady signal which becomes, on average, stronger at high
observing frequencies. 
There is no obvious correlation found between $P_3$ and
the pulsar age (or any other pulsar parameter) contrary to reports in the past. 
\end{abstract}

\maketitle

\section{Introduction}

\begin{figure}[tb]
\rotatebox{270}{\resizebox{1.4\hsize}{!}{\includegraphics[angle=0]{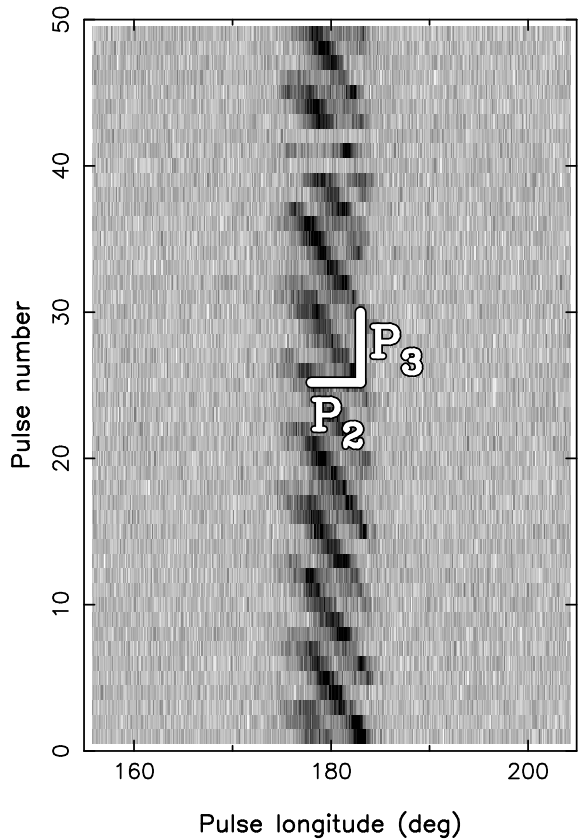}}}
\caption{\label{stack}An example of a pulse-stack of
fifty successive pulses of PSR B0818--13 as observed by the WSRT at 92 cm. Two successive drift bands
are vertically separated by $P_3$ and horizontally by $P_2$.}
\end{figure}

Although the pulse profiles of radio pulsars are in general very
stable, the shape of their single pulses are often highly variable
from pulse to pulse. For some pulsars the single pulses are modulated
in a highly organized and fascinating way: they exhibit the phenomenon
of drifting subpulses. An example is shown in
the left panel of Fig. \ref{stack}. In this so-called
``pulse-stack'' fifty successive pulses are displayed on top of one
another and a beautiful pattern of diagonal ``drift bands'' emerges.

There are a few types of models that attempt to explain the drifting
phenomenon. The most well known model is the carousel model
(\citet{rs75}), which has been extended by many authors
(e.g. \citet{gmg03}) making it the most
developed model for explaining the drifting phenomenon.  These models
explain the drifting phenomenon by the generation of the radio
emission via a rotating ``carousel'' of discharges which circulate
around the magnetic axis due to an $\mathbf{E}\times\mathbf{B}$
drift. Alternative models to explain drifting subpulses are the model for non-radial pulsations of neutron stars (e.g. \citet{cr04}) and the feedback model proposed by \citet{wri03} . 

We have embarked on an extensive observational program to survey a
large sample of pulsars to study their single pulse modulation using
the Westerbork Synthesis Radio Telescope (WSRT) in the
Netherlands. The main goals of this program are to determine what
fraction of the pulsars have drifting subpulses, whether those pulsars
share some physical properties and if subpulse modulation is frequency
dependent. The sample of pulsars studied is selected based only on the
predicted $S/N$ in a reasonable observing time, which makes the
resulting statistics as unbiased as possible towards well-studied
pulsars, pulse profile morphology or any particular pulsar
characteristics. If possible, we observed the pulsars at two
wavelengths around 21 and 92 cm.

The results of the pulsars observed at a wavelength of 21 and 92 cm are
published in \citet{wes06} and \citet{wse07} respectively. All the
plots of the two 21-cm and 92-cm observations can also be found side by
side in the PhD. thesis of the main author.\footnote{The thesis of
P. Weltevrede is online available via the following \url {http://dare.uva.nl/en/record/217315}, or contact the author for a hardcopy.}
In this proceedings we summarize the results of the observations and a
comparison between the two observing frequencies is made.

\section{Results}
\subsection{Drifting subpulses are very common}

Our sample of pulsar is not biased on pulsar type or any particular
pulsar characteristics. This allows us, first of all, to address the
very basic question: what fraction of the pulsars show the drifting
phenomenon?  Of the 187 analyzed pulsars at 21 cm, 68 pulsars
show the drifting phenomenon. At 92 cm this fraction is a little bit higher: for 76 of the 185 analysed pulsars we found drifting subpulses. Most of the pulsars for which we detected drifting subpulses
were not known to have them. This shows first of all that the used method to detect drifting subpulses (using fluctuation spectra; \citet{es02}) works extremely well.

For about one in three
pulsars we found drifting subpulses, and this is a lower limit for a number of reasons. The most important 
reason is that the chance of finding drifting subpulses
was found to be correlated with the $S/N$ ratio of the observation. The probability of detecting drifting is higher for
observations with a higher $S/N$.  We estimate that at least half of the pulsars have drifting subpulses.
There are many reasons why drifting is not expected to be
detected for all pulsars. 
For instance for some pulsars the line of
sight cuts the magnetic pole centrally and therefore longitude
stationary subpulse modulation is expected. Also, refractive
distortion in the pulsar magnetosphere or nulling (periods during which the pulsar emits no radio emission) will disrupt the
drift bands, making it difficult or even impossible to detect
drifting. Some pulsars are known to show organized drifting subpulses
in bursts. In that case (or when $P_3$ is very large) some of our
observations could be too short to detect the drifting.

With a lower limit of one in two it is clear that drifting is at the
very least a common phenomenon for radio pulsars and it could well
be that the drifting phenomenon is an intrinsic property of the
emission mechanism although for some pulsars it is difficult or even
impossible to detect.

\subsection{Drifting subpulses become more organized when the pulsar ages}

\begin{figure*}[t]
\rotatebox{270}{\resizebox{!}{0.95\hsize}{\includegraphics[angle=0]{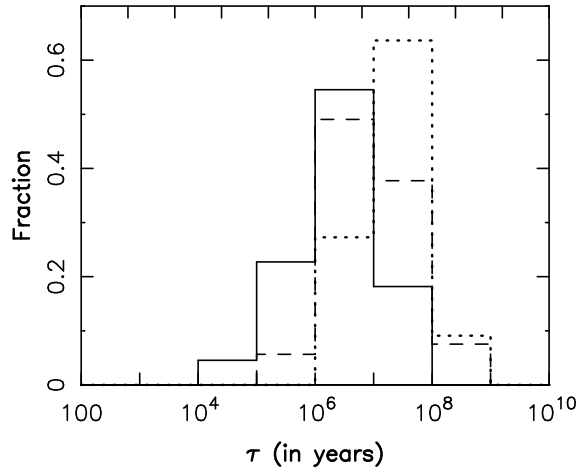}}}
\caption{\label{age_hist92}The histogram of the
characteristic ages of the analyzed pulsars at 92 cm with a $S/N$ $\geq100$.  The
solid line is the age distribution of pulsars without drifting
subpulses, the dashed line shows all the drifters and the dotted line
shows the coherent drifters.}
\end{figure*}

In Fig. \ref{age_hist92} it can be
seen that the population of pulsars that show the drifting phenomenon
is on average older than the population of pulsars that do
not show drifting.
Moreover it seems that the population of pulsars that show coherent
drifting (i.e. very \emph{regular} drifting subpulses) is on average older than those who show
less regular drifting subpulses. The same trend is found both in the 21 and the 92 cm data.
It is intriguing to think that drifting
becomes more and more coherent for pulsars with a higher age. This 
correlation cannot be explained by nulling, because the nulling fraction is on average higher for
older pulsars. Possibly the alignment of the magnetic dipole
axis with the rotation axis has something to do with the observed
trend. Observations seem to show that the angle $\alpha$ between the
magnetic axis and the rotation axis is on average smaller for older
pulsars and this angle is likely to be an important physical parameter
in the mechanism that drives the drifting phenomenon.  In this
scenario as the pulsar gets older, the rotation axis and the magnetic
axis grows more aligned, which makes the drifting mechanism more
effective or regular. 

\subsection{Subpulse modulation at the two frequencies}

An interesting quantity to consider is the modulation index, which is a measure of the factor by
which the intensity varies from pulse to pulse. It is clear that the modulation index is a parameter that is closely
related with the drifting phenomenon, because drifting subpulses imply
an intensity modulation. However, it is somewhat arbitrary how {\em
the} modulation should be defined, because the longitude-resolved
modulation index is in most cases highly dependent on pulse
longitude. We have used the minimum in the modulation index profile, which should be more independent of the $S/N$ of the observation.

\begin{figure}[tb]
\rotatebox{270}{\resizebox{!}{0.94\hsize}{\includegraphics[angle=0]{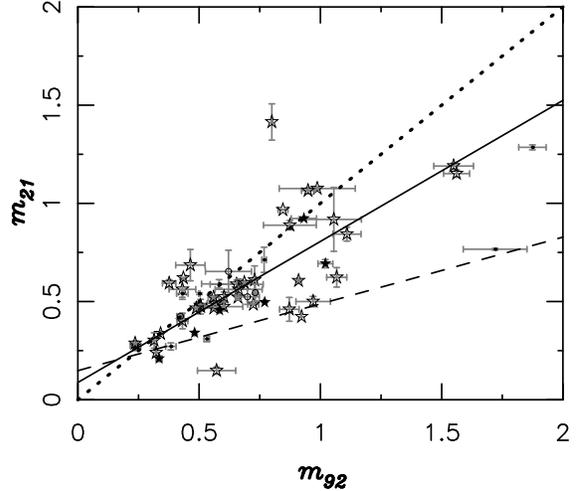}}}
\caption{\label{mod_mod}The modulation index measured at 21 cm versus
that measured at 92 cm. The points would lie on the dotted line when
the modulation index is frequency independent. Only observations with
a $S/N\geq 100$ are included this plot. The dashed and solid line are
straight line fits through the data-points with and without weighting
for the errorbars. The pulsars without drifting
subpulses are the dots, the pulsars with drifting subpulses are
the open stars, the pulsars with very regular drifting subpulses are the filled stars and the
pulsars showing longitude stationary subpulse modulation are the open
circles.}
\end{figure}

In Fig. \ref{mod_mod} it can be seen that the modulation indices
measured at the two frequencies are not independent, but they are
clearly correlated.  If the modulation index was independent of
observing frequency the points would be scattered symmetrically around
the dotted line.  However, there are more pulsars with
a higher modulation index at 92 cm than visa versa. 

This trend is also confirmed by making a straight line
fit through the data. This is done by minimizing the $\chi^2$ incorporating the
measurement errors of both coordinates (dashed line). To avoid the best fit
being dominated by a few high $S/N$ observations, the fitting is also
done by weighting the data-points equally. This fit also confirms that the
modulation index at 92 cm is on average higher than at 21 cm (solid
line).

\subsection{Properties of drift behavior}

\begin{figure}[tb]
\rotatebox{270}{\resizebox{!}{0.94\hsize}{\includegraphics[angle=0]{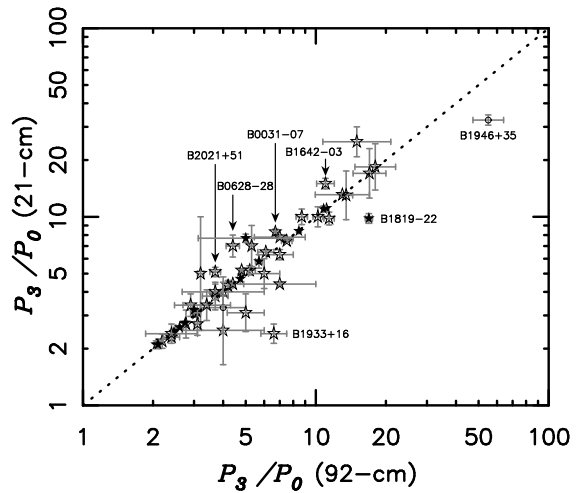}}}
\caption{\label{p3p3} The measured value of the vertical drift band
separation $P_3$ compared at the two frequencies. Also the low-$S/N$
observations are included. The dotted line shows the expected correlation when $P_3$ is
independent of frequency. See the caption for Fig. \ref{mod_mod} for the meaning of the different symbols.}
\end{figure}

It is generally thought that the value of $P_3$ is independent of the
observing frequency, while the value of $P_2$ could vary. All pulsars with a measured $P_3$ at two
frequencies are compared in Fig. \ref{p3p3} to test  the
absence of a dependency on the observing frequency of $P_3$. The
correlation is indeed  extremely tight. This correlation confirms the
report for nine pulsars by \citet{nuwk82}.  Moreover, many points that
do not fall on the  correlation can be explained. In
the case of PSRs B0031$-$07 and B1819$-$22 drift-modes with different
$P_3$ values dominate at the two frequencies. For others, such as PSRs
B1738$-$08 and B1946+35 it seems  not unlikely that
longer observations will reveal that the $P_3$ values are consistent
at the two frequencies.

There is no obvious correlation found between $P_3$ and
the pulsar age contrary to reports in the past. The absence of a
correlation with age in this large sample of pulsars with drifting
subpulses (90) suggests that if such a
correlation would exist, it must be a very weak correlation. Also, the
evidence for a pulsar sub-population located close to the $P_3=2P_0$
Nyquist limit (\citet{wri03,ran86}) is weak.

\section{The quasi-steady emission component}

\citet{nuwk82} concluded, based on a multi-frequency study of nine
pulsars, that it is a common feature of pulsars to have a quasi-steady
component in their emission which becomes stronger with increasing
frequency. Our observations support this conclusion in several ways.
First of all, a considerable fraction of the pulsars with drifting
subpulses have a modulation index lower than $m = 1/\sqrt{2}$ (see Fig. \ref{mod_mod}). Such low modulation indices cannot be produced by a  pure drifting subpulse signal. Fig. \ref{mod_mod} also shows the trend that the modulation index
is, on average, lower at 21 cm. This would be consistent with the idea that
there is a quasi-steady component in the emission of pulsars which is
relatively strong at higher frequencies. 
This picture may also explain why the chance of detecting drifting
subpulses is slightly higher at low frequencies, because
the drifting subpulse
signal is relatively stronger at lower frequencies. 

\section{Summary and Conclusions}

Drifting subpulses are at
the very least a common phenomenon for radio pulsars, if not an
intrinsic property of the emission mechanism. For
57 pulsars drifting subpulses are discovered for the
first time, showing the success of this survey. It
is estimated that at least half of the total population of pulsars
will show drifting subpulses when observations with high enough $S/N$
would be available. The chance of detecting drifting subpulses
at both frequencies is high, indicating that the drifting phenomenon
is in general broadband. 

Drifting subpulses are at the very least a common
phenomenon for radio pulsars, which implies that the physical
conditions required for the emission mechanism and the drifting
mechanism to work are similar. This is consistent with the absence of
a correlation with the surface magnetic field strength. 

The measured values of $P_3$ at the two frequencies are
highly correlated. These correlations are
expected when the drifting subpulses share a common physical origin.
A correlation between $P_3$ and other pulsar parameters is expected if 
the drift rate depends on any physical parameters of the pulsar and the
strongest correlation is expected to be found when $P_2$ is
 identical for different pulsars. Such a correlation
would be a very important observational restriction on pulsar emission
models. However, there are no such correlations found. This could suggest that many pulsars in our sample are aliased
or that $P_2$ is highly variable from pulsars to pulsar.

Our sample of pulsars is not biased on pulsar type or any particular
pulsar characteristics, which allows us to do meaningful statistics on
the drifting phenomenon. There is a weak trend found that pulsars with
drifting subpulses are on average older, especially the very regular
drifters. This correlation
suggests that there is an evolutionary trend such that the youngest
pulsars have the most disordered subpulses and that the subpulses
become more and more organized into drifting subpulses when the pulsar
ages. This trend could for instance be explained by the evolution of the angle
between the magnetic axis and the rotation axis or the evolution of
the pulse morphology. In the non-radial pulsations model
(\citet{cr04}) this trend can also be explained, because the
appearance of narrow drifting subpulses is favored in pulsars with an
aligned magnetic axis.  This trend cannot
explained by nulling. 

The modulation indices measured at the two frequencies are clearly
correlated, although they tend to be higher at low
frequencies.  This is consistent with the picture in
which the radio emission can be divided into a drifting subpulse
signal plus a  quasi-steady component which becomes
stronger at high observing frequencies.


\bibliographystyle{aipproc}   

\bibliography{talk}
\end{document}